\documentclass[pdflatex,sn-mathphys-ay]{sn-jnl}%
\usepackage{graphicx}%
\usepackage{multirow}%
\usepackage{amsmath,amssymb,amsfonts}%
\usepackage{amsthm}%
\usepackage{mathrsfs}%
\usepackage[title]{appendix}%
\usepackage{xcolor}%
\usepackage{textcomp}%
\usepackage{manyfoot}%
\usepackage{booktabs}%
\usepackage{algorithm}%
\usepackage{algorithmicx}%
\usepackage{algpseudocode}%
\usepackage{listings}%

\usepackage{bbm}
\usepackage{xcolor}
\usepackage{booktabs}

\usepackage[normalem]{ulem}
\newcounter{question}
\newcommand{\question}[1]{%
  \refstepcounter{question}%
  \noindent\textbf{Question \thequestion.}\label{q:\thequestion}~ #1
}

\usepackage{bbm}
\usepackage{booktabs}
\usepackage{adjustbox}
\usepackage{comment}
\usepackage{hyperref}

\usepackage[english]{babel}

\theoremstyle{thmstyleone}%

\theoremstyle{thmstyletwo}%

\theoremstyle{thmstylethree}%

\begin{document}

\title[yeah]{The Perceived Impact of Environment on Health in Italy: a Penalized Ordinal Regression Approach}

\author*[1]{\fnm{Mattia} \sur{Stival}}\email{mattia.stival@unive.it}
\author[1]{\fnm{Angela} \sur{Andreella}}
\author[1]{\fnm{Gaia} \sur{Bertarelli}}
\author[1]{\fnm{Catarina} \sur{Midões}}
\author[1]{\fnm{Stefano Federico} \sur{Tonellato}}
\author[1]{\fnm{Stefano} \sur{Campostrini}}

\affil[1]{\orgdiv{Department of Economics}, \orgname{Ca' Foscari University of Venice}, \orgaddress{\street{San Giobbe, Cannaregio 873}, \city{Venice}, \postcode{30121}, \state{Italy}}}

\abstract{
Understanding how individuals perceive their living environment is complex, as it reflects both personal and contextual determinants. 
Yet, understanding environmental perceptions is relevant to better address and target health and sustainability policies.
In this context, data analyses are challenged by the need to combine large-scale individual survey data with detailed contextual indicators,
while respecting the ordinal nature of perception measures, and accommodating the presence of high-dimensional, correlated covariates.
We resolve these challenges by employing a penalized semi-parallel cumulative ordinal regression model. 
The model automatically handles both parallel and non-parallel effects and employs regularization to address well-known issues, such as multicollinearity, separation, and covariate selection.
The approach is applied to data from the environmental module of the Italian health surveillance system, PASSI, integrated with municipal-level contextual information, including socio-economic indicators, pollution exposure, and geographical characteristics.
 The results reveal substantial territorial heterogeneity in environmental perceptions and show how both individual-level and local contextual factors, including hazardous environmental conditions, shape subjective evaluations.
Overall, the analysis illustrates the usefulness of the semi-parallel ordinal framework for providing relevant insights into environmental and public health policy design.}

\maketitle

\section{Introduction}

Understanding how citizens perceive the environment in which they live is fundamental, as perceptions are closely related to individual and collective well-being. 
Not only do they reflect the quality of life in a given context, but they also shape awareness of risks and opportunities, guide daily choices that affect health, and inform the design of policies and communication strategies in the domains of health promotion, environmental protection, and social equity.
.

Despite its relevance, as recognized in early works by \cite{buttel1976environmental, van1981environmental, dunlap1984commitment, mohai1987age}, environmental perception remains incompletely understood. Environmental perception emerges from the joint contributions of individual characteristics (e.g., age, gender, education, socioeconomic status, psychological traits, political orientation) and place-based factors (e.g., urban form, natural resource quality, pollutant exposure, territorial vulnerabilities). 
The very nature of the problem entails several sources of complexity. 
On one side, it requires data that explicitly captures individuals’ perceptions of their surrounding environment. 
Large-scale surveys devoted to these dimensions remain limited, typically conducted by national statistical institutes. As a result, much of the existing evidence is based on small-scale or local studies \citep[][]{coi2016risk, MARTINKERRY202385}.
On the other side, even when such surveys are available \citep[see, for the Italian case, ][]{sampaolo2017perception, ISTAT_env}, they often provide rich information at the individual level but
limited detail on the contextual conditions in which respondents live. This limits the analysis of how
environmental perceptions relate to local environmental and territorial characteristics.

In the literature, only a limited number of studies attempt to address these limitations.  \cite{salvatore2024use} apply small area estimation techniques to obtain reliable estimates of the proportion of people who are very and extremely worried about climate change at regional level, incorporating non-traditional auxiliary information, such as web data from the  \cite{ess2022}. 
\cite{hannibal2016personal} use data from nationally representative surveys and government agency data 
to examine the extent to which local environmental stressors, such as air pollution and industrial waste, are related to individual environmental concern.
Within this context, data integration methods \citep[e.g.][]{yang2020statistical, kim2021data} play a pivotal role, enabling researchers to combine complementary sources into a coherent analytical framework. 
An example is provided by \cite{midoes2024share},  who construct the SHARE-ENV dataset by integrating SHARE (Survey of Health, Ageing and Retirement in Europe) data \citep{share_release9_2022}, with environmental information to study the interaction between climate change, environmental stressors, and vulnerability factors.

Despite their potential, analyzing these integrated data sources remains a methodological challenge, which highlights the need for statistical models designed to study environmental perception, particularly in data integration settings.
On the one hand, statistical models must accommodate the structure of survey outcomes and covariates, including ordinal response variables and potentially high-dimensional covariate spaces, given the large number of individual and contextual indicators that may be relevant to include in the analysis.
On the other hand, the resulting models should provide interpretable outputs. For instance, standard black-box approaches to variable selection and modeling, while effective, may lack interpretability to support policy-relevant conclusions.

To the best of our knowledge, most existing studies either rely on data that are insufficiently rich to support such analyses or adopt relatively simple modeling strategies. 
While such studies generally rely on regression frameworks, our work emphasizes that penalization strategies are critical for addressing data integration challenges in studies of environmental perceptions, particularly when ordinal responses are used.
Ordinal variables provide a natural and interpretable way to represent attitudes and opinions, as well as latent constructs such as perceptions and preferences \citep{mccullagh1980regression, agresti2010analysis}.
Penalization is a well-established regularization strategy that promotes parsimony and supports stable estimation and variable selection in high-dimensional settings \citep{tibshirani_lasso_1996, elasticnetZou}. Although penalization is widely used in ``large $p$–small $n$'' problems \citep{waldron2011optimized, frank1993statistical}, its integration with flexible ordinal regression models remains limited in the empirical studies of environmental perceptions.

Recently, \cite{JSSordinalElNet} introduced a penalization approach for elementwise link multinomial–ordinal (ELMO) models within the class of vector generalized linear models \citep{yee2015vector}. 
Building on this framework, we address the data integration challenges discussed above by adopting a semi-parallel ordinal specification, which allows the model to flexibly move between proportional-odds (parallel) and non-proportional-odds (non-parallel) assumptions. 
While parallel models are commonly used due to their simplicity, their restrictive assumptions may be unrealistic in large-scale analyses of environmental perceptions over a nationwide territory.
Allowing for non-parallel effects enables the model to capture heterogeneity in environmental awareness, i.e., differences in recognition, understanding, and knowledge of environmental issues \citep{PEREA2025e43679, monus2021environmental}.

We apply this framework to the Italian case by exploiting the richness of micro-level information provided by the PASSI system (Progressi delle Aziende Sanitarie per la Salute in Italia), a repeated cross-sectional behavioral risk-factor surveillance survey conducted by the Italian National Health Institute.
Using a conditional ordinal variable with three categories, 
we study how Italian citizens perceive the impact on health of the environment in which they live.
Given the considerable heterogeneity of the Italian territory, in terms of population composition, geographical conditions, wealth, risk exposure, and lifestyles \citep{istat2023bes,istat_poverty_2024, oecd_hows_life_2024}, 
we incorporate municipal-level contextual variables from different sources to enrich individual information available in PASSI.
More specifically,  we considered PM2.5 concentration
derived from Copernicus Atmosphere Monitoring Service (CAMS) global reanalysis \citep[EAC4,][]{inness2019cams}, a set of municipal indicators from ISTAT, such as the municipal population density, their geographical location (mountain or coastal), and other measures of socio-economic and environmental vulnerability.
Our aim is to disentangle how the perceived effects of the environment on personal health relate simultaneously to individual features and to broader contextual characteristics.
We describe how such models can be used to explore two fundamental dimensions of perception: 
the determinants of \emph{positivity} and of \emph{neutrality}.
The first reflects characteristics that, if positive (or negative), lead to a positive (or negative) attitude toward the surrounding environment. Positivity indicates satisfaction and adaptation, whereas its negative counterpart points to underlying concerns.
The second captures those characteristics that are mostly associated with increased neutrality. 
Departing from neutrality may represent an initial step toward heightened awareness and active engagement.
Clarifying these relationships is important for identifying the drivers of perceived inequalities and for informing policies that can effectively address both personal vulnerabilities and contextual disadvantages, enhancing the understanding of the interplay among environmental beliefs, health, and awareness. 

The paper is organized as follows. Section~\ref{sec:data} introduces the PASSI dataset and its environmental module, together with the complementary data sources used in the analysis; Section~\ref{sec:method} describes the proposed methodological framework; Section~\ref{sec:res} reports the results and offers a critical discussion; and Section~\ref{sec:conclusion} concludes with implications and directions for future research.
The Supplementary Material (SM) contains further analyses and insights regarding the study.

\section{Background and data}
\label{sec:data}

In this section, we introduce the data in Subsection~\ref{subsec:data} and present exploratory analyses in Subsection~\ref{subsec:eda}.

\subsection{The PASSI surveillance system and the external datasets}\label{subsec:data}

PASSI is a nationwide surveillance system that, since 2008, has continuously collected data on behavioral risk factors and health among Italian adults aged 18-69 years \citep{baldissera2011peer}. 
Based on the US Behavioral Risk Factor Surveillance System, the PASSI surveillance system aims to establish a continuously updated, local-level database to monitor trends in health issues, risk factors, and preventive measures in Italy. 
The system adopts a stratified sampling design by gender and age at the Local Health Unit (LHU) level (ASL, Azienda Sanitaria Locale), based on the health registries. The gender-age specific strata must include at least six categories (i.e., men aged 18-34, men aged 35-49, men aged 50-69; women aged 18-34, women aged 35-49, women aged 50-69). 
In this way, the sample is representative and the survey provides reliable estimates at the LHU-age-sex level.
Excluding Lombardy, all regions participate in the survey.
Every year, over 90\% of LHUs operating across Italy is in the surveillance, providing information on over 90\% of the country’s resident population, with a response rate consistently above 85\%, while the refusal rate never exceeds 10\% \citep{iss_2023_protocollo_passi}.

In 2023, the total sample size is $31532$.
Due to differences in population size across regions, sample sizes vary accordingly, from $244$ respondents in Molise to $3546$ in Emilia Romagna.
On behalf of the LHUs, data are collected via a telephone survey administered by trained staff. 
The survey
is composed of
a series of predefined modules with questions on different health-related topics, such as chronic disease prevention, lifestyle behaviors, screening participation, mental health, and social determinants.  
Being a rich source of information, PASSI data constitute a well-established resource for researchers in several fields in Italy, e.g., for studies on morbidity \citep{Pastore_passi22,andreella2023,stival2024bayesian}, health in urban areas \citep{nobile2022urban}, and behavioral profiles \citep{minardi2011social, minardi2022p06, gorini2023electronic, andreella2023}. The system also includes modules on emerging public health priorities.
Among them, module 18 investigates how individuals perceive the environment in relation to their personal health.
However, because this module was optional until 2023 and not consistently implemented across regions, studies concerning these aspects are few and dated \citep{sampaolo2017perception}.
 
Referring to the PASSI 2023 survey, in this work, we consider the first item of the environmental module, which is: \vspace{1mm}\\
\question{\label{q_1}
{Thinking about the environment of your neighborhood or the area where you live, how do you think it influences your current state of health?}}
\vspace{1mm}
\\
{
Respondents could choose among four options: 
\emph{i}) it does not; 
\emph{ii}) it does, positively; 
\emph{iii}) it does, negatively; 
\emph{iv}) I don't know.
We focus attention on the analysis of subjects who chose the first three options (corresponding to more than 95\% of the sample).
Thus, we model the \emph{respondent’s belief about the environment’s influence on their health} as being either \emph{positive} (if \emph{ii}), \emph{neutral} (if \emph{i}), or \emph{negative} (if \emph{iii}), conditionally on not having chosen the \emph{I don't know} option.
The exclusive and separate treatment allows us to study the response variable of interest as being a (conditionally) ordinal variable, in which  \emph{I don't know} respondents are not mixed with other modalities.
This modality does not represent an environmental perception (not even a neutral one). Therefore, we present a separate analysis and further motivations in the SM.
Considering this item is particularly appealing for studying how respondents express their beliefs regarding the environment, not only in terms of worries but also with respect to potential positive aspects of their living surroundings. 
}
Our interest lies in analyzing variation in the response by jointly considering individual and contextual data, detailed in Table~\ref{tab:variables}.

Individual data come from the PASSI dataset and include socio-demographic variables (age, sex, occupation, citizenship, educational level, economic status, and co-living situation), spatial information (the respondent’s Local Health Unit), and a health indicator (reporting at least one chronic disease).
In addition, we incorporate municipal-level contextual variables coming from the SHARE-ENV \citep{midoes2024share} and ISTAT (Italian National Institute of Statistics) databases. 
These include the yearly average of PM2.5 concentration by municipality, derived from Copernicus Atmosphere Monitoring Service (CAMS) global reanalysis \citep[EAC4,][]{inness2019cams} as in the work of \cite{midoes2024share}, municipal population density, municipal location  \citep[mountain, coastal, ][]{istat_2023_ecoregioni}, and the municipal fragility index \citep[MFI,][]{istat_municipal_fragility_index_2023}. 
The MFI index provides a composite measure of socio-economic and environmental vulnerability at the municipality level, capturing aspects such as demographic aging, economic hardship, environmental exposure, and limited access to essential services. 
More details are available at \url{https://www.istat.it/en/press-release/municipal-fragility-index-ifc/}.

\begin{table}[ht]
\centering
\begin{adjustbox}{max width=\textwidth}
\begin{tabular}{|l|l|p{7cm}|l|}
\toprule
\textbf{Variable} & \textbf{Type} & \textbf{Description} & \textbf{Source} \\
\midrule
1. Local Health Unit (LHU)       & Categorical ($103$ levels) & Belonging of the respondent to the LHU (stratification variable) & PASSI\\
2. Age Class          & Categorical ($3$ levels) & Age groups (in years): $[18,34]$, $[35,49]$, $[50, 69]$ (stratification variable) & PASSI \\
3. Sex & Binary & Sex assigned at birth: Male/Female (stratification variable) &  PASSI\\
\midrule 
4. Chronicity          & Binary & Yes if the respondent has at least one chronic disease (diabetes, kidney failure, respiratory diseases, cardiovascular diseases, and tumors) & PASSI   \\
5. Shared living    & Binary & Living with others (e.g. family) &PASSI   \\
6. Occupation        & Binary & Yes if employed at the date of the interview &PASSI   \\
7. Citizenship         & Binary & Yes if the respondent declares to have the Italian citizenship &PASSI   \\
8. Educational level         & Binary & High if the respondent has completed at least a high school diploma (Diploma Superiore) &PASSI   \\
9. Economic level         & Binary & High if the respondent reports no economic difficulty & PASSI   \\
\midrule
10. Fragility index  & Numerical & 
Municipal Fragility index 
& ISTAT   \\
11. PM2.5    & Numerical & Yearly average PM2.5 at the municipality level & SHARE-ENV  \\
12. Log-population density       & Numerical & Logarithm of population density at the municipality level & ISTAT \\
14. Mountain  & Binary & Yes if municipality has center at least  $600$ m above the sea & ISTAT\\
15. Coastal & Binary & Yes if the municipality is in a coastal zone & ISTAT\\
\bottomrule
\end{tabular}
\end{adjustbox}
\caption{Description of variables used in the analysis}
\label{tab:variables}
\end{table}

\subsection{Exploratory data analysis}\label{subsec:eda}
\begin{figure}[t]
    \centering
    \includegraphics[width=0.99\linewidth]{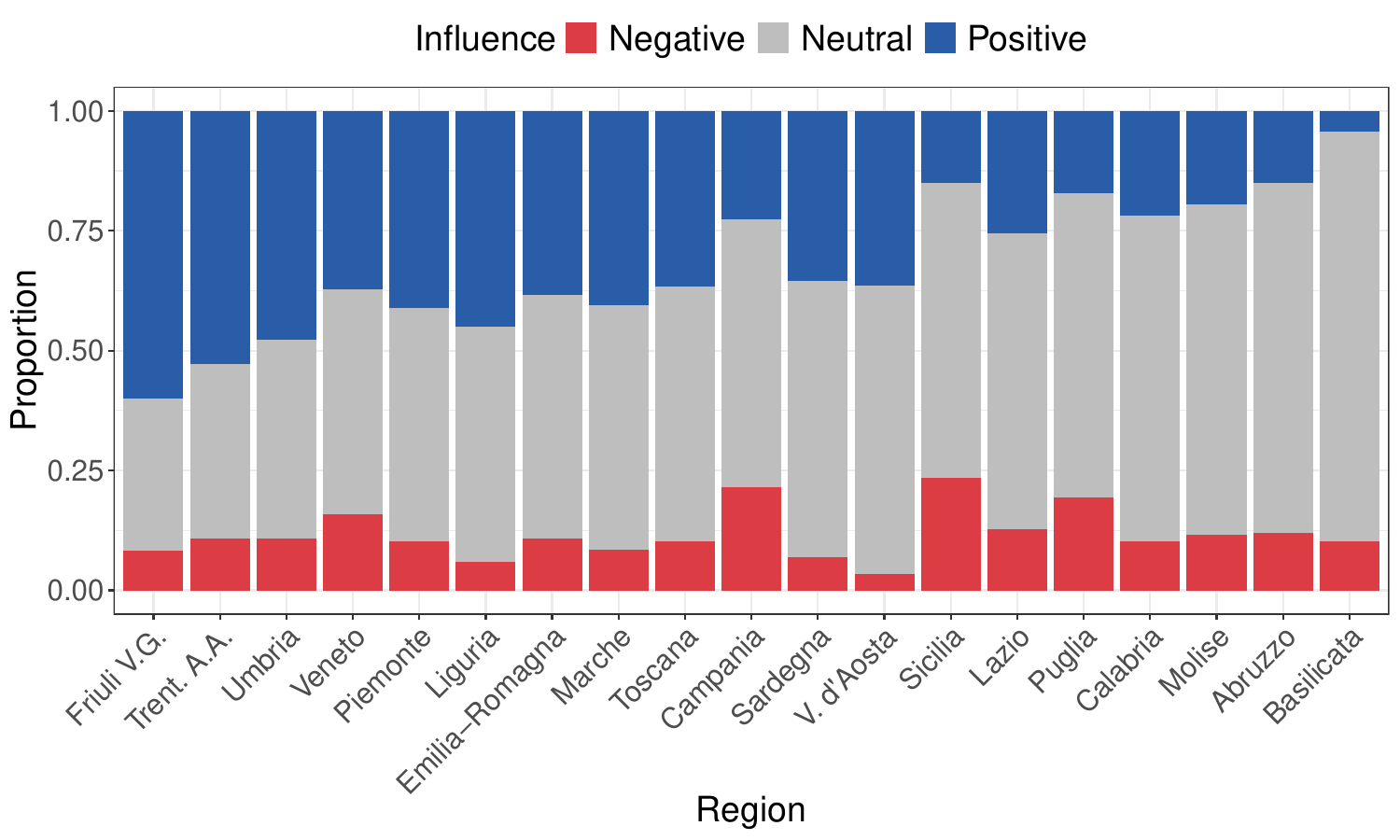}        
    \caption{
    Perceived environmental health influence proportion by Region, ordered by the proportion reporting 'No' (neutral) influence.    }
    \label{fig:regional_influence}
\end{figure}

\begin{figure}[t]
    \centering
    \includegraphics[width=0.89\linewidth]{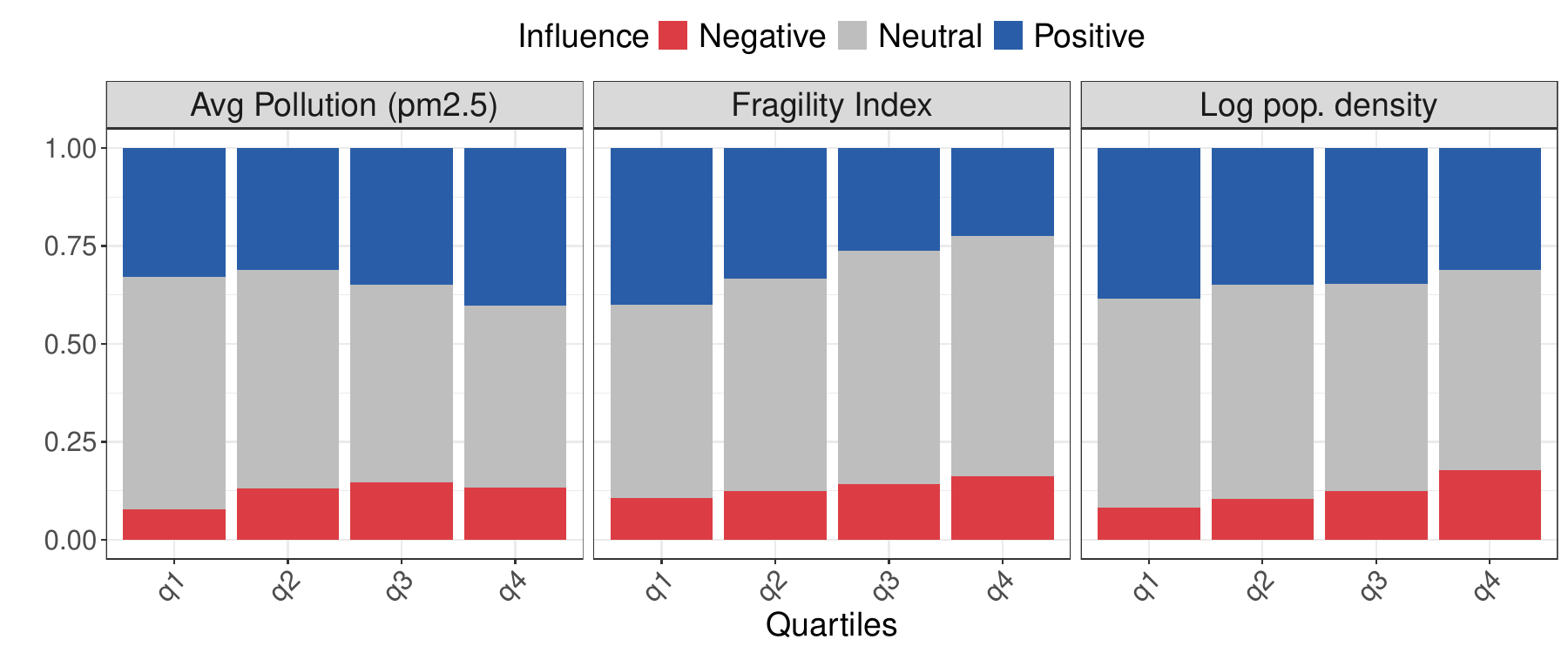}
    \includegraphics[width=0.89\linewidth]{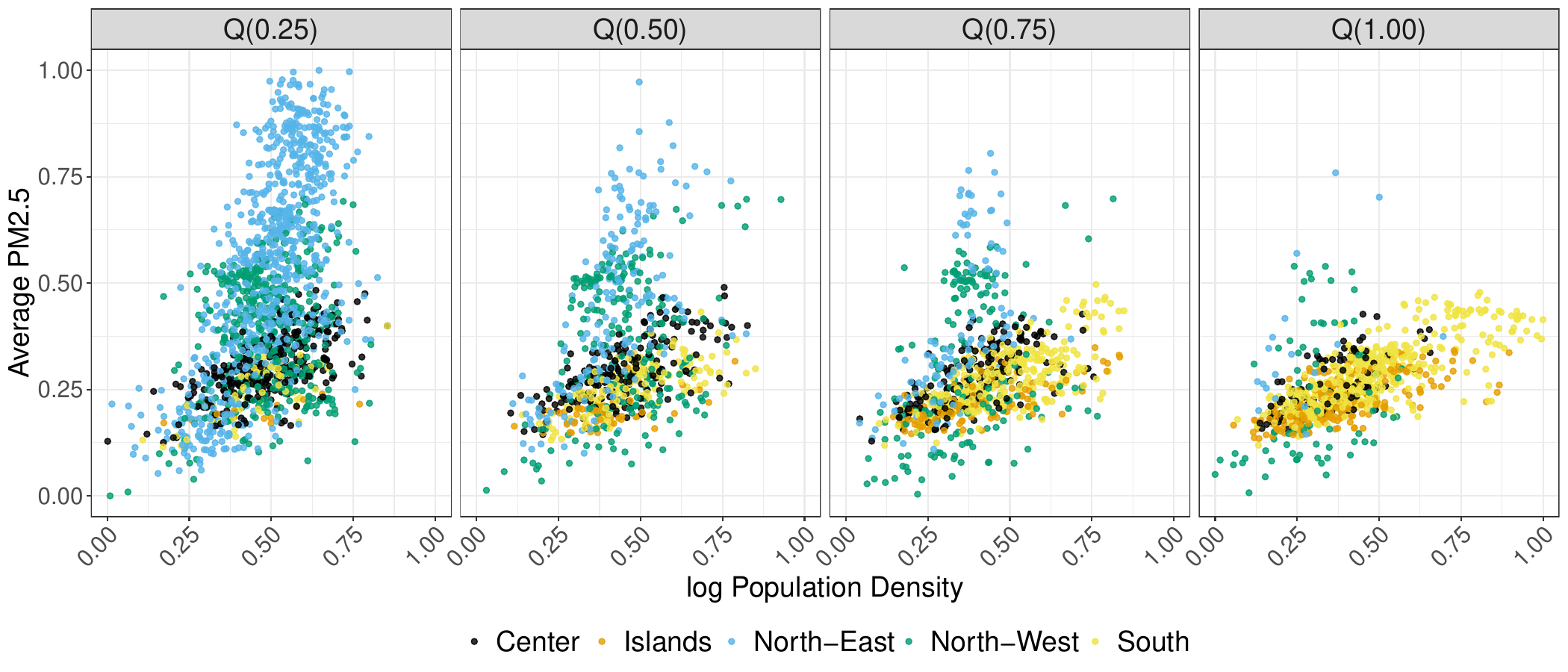}
    \caption{The top panel illustrates the variation in responses across quartile classes of contextual variables measured at the municipal level: average pollution, the FMI, and log-transformed population density. 
    The bottom panel displays a scatter plot of log-population density against the average PM2.5 level for each municipality, stratified by quartile classes of the MFI. Points are colored according to the municipality's macro-area. 
    For average pollution and MFI, higher quantiles indicate worse conditions. 
    }

    \label{fig:ISTAT}
\end{figure}

We present exploratory plots to assess whether perception is associated with both contextual and individual factors in Italy. 
Figure~\ref{fig:regional_influence} shows the distribution of responses across Italian regions, ordered by the proportion of neutral respondents.
Substantial between-region variability emerges, mainly in the proportion of neutral and positive evaluations.
For instance, Friuli-Venezia Giulia exhibits both the lowest share of neutral responses and the highest share of positive ones. 
In contrast, Basilicata exhibits the highest proportion of neutral responses and the lowest proportion of positive evaluations.
This between-region variability likely reflects the pronounced heterogeneity of the Italian context, encompassing differences on physical geography, population composition, socio-economic conditions, wealth, exposure to environmental risks, lifestyles, and health status \citep{stival2024bayesian}.
For example, approximately $30\%$ of municipalities in Friuli-Venezia Giulia have an MFI above the national average,
compared to $86\%$ in Basilicata.
To this end, Figure \ref{fig:ISTAT} explores the association between municipality-level variables, obtained from external data sources (i.e., SHARE-ENV and ISTAT), and the responses to Question~\ref{q_1}.

The upper panel presents the distribution of responses to Question~\ref{q_1} across classes defined by quartiles of average pollution (PM 2.5 concentration), the MFI, and log-transformed population density.
We observe that, in the upper quantiles of both the MFI and log-population density, the share of respondents reporting a negative impact of the environment on their health increases, whereas the share reporting a positive impact correspondingly decreases.
This essentially corresponds to a deterioration in the perception as contextual conditions in which people live tend to worsen.
Notably, this trend is less evident for pollution (first plot in the upper panel).
The bottom panel, which reports scatter plots of average pollution levels versus log-population density by municipality--scaled using the min-max transformation--for different quartile classes of the MFI, can partially explain this apparent inconsistency. 

\begin{figure}
    \centering
    \includegraphics[width=0.5\linewidth]{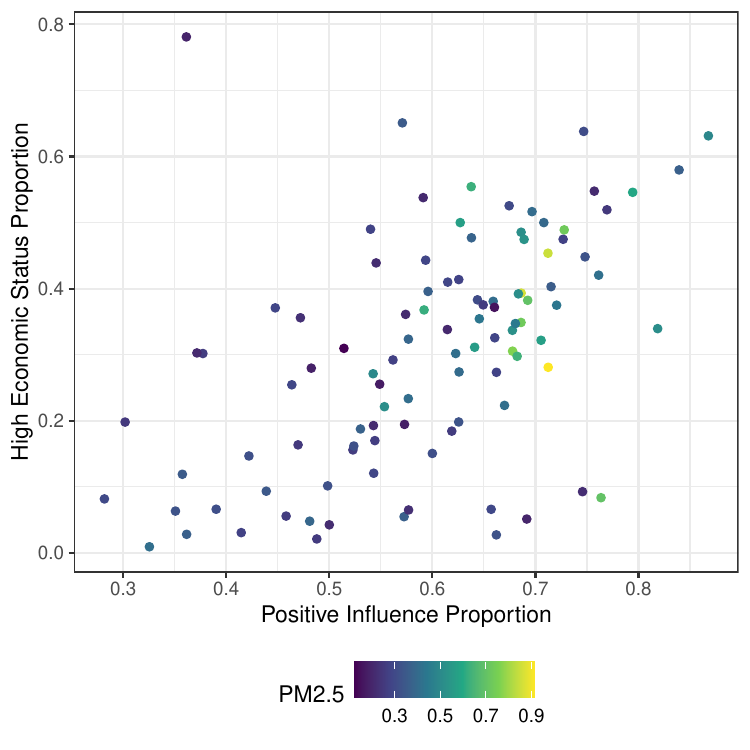}
    \caption{Proportion of respondents by LHU reporting a positive environmental impact on health and no difficulty making ends meet. Colour gradient encodes the mean LHU pollution level.}
    \label{fig:asl}
\end{figure}

Notably, municipalities with high pollution levels (left plot) often have a low MFI.
These municipalities are mainly located in the North-East and North-West macro areas of Italy, which are known for their strong industrial bases. 
While this industrial activity correlates with higher pollution levels, these areas also have higher per capita income and overall socioeconomic development, resulting in lower MFI, better economic conditions, and other correlated aspects that may also improve environmental perception.
On the one hand, this implies that these indicators alone are insufficient to capture the full range of dynamics present in the data; on the other hand, it is important to recognize that these factors are correlated and linked to the population characteristics of different geographical areas. 
This last aspect can be understood by looking at the data from different perspectives. For instance, Figure \ref{fig:asl} plots the proportion of individuals reporting a positive environmental influence against the proportion of individuals with high economic status, aggregated by LHU. The color gradient indicates the average PM2.5 concentration in each LHU, computed considering the arithmetic mean by LHU.
The positive relationship between high economic status and positive perception of environmental influence is apparent. However, examining PM2.5 concentrations, we observe several LHUs characterized by both high PM2.5 levels and a high proportion of individuals reporting a positive influence. 

Since economic resources can mitigate the perceived negative effects of environmental exposure, a simultaneous approach to addressing both these factors is necessary.
In so doing, we must respect the ordinal nature of the response. 
Accordingly, the following section introduces a regression model for ordinal responses with semi-parallel effects, initially proposed by \cite{JSSordinalElNet}. 
As explained in more detail below, the framework is flexible enough to include numerous covariates, consider potential interactions, and depart from the typical assumption of parallel odds while maintaining interpretability.
For completeness, additional exploratory results are reported in the SM.

\section{Methods}\label{sec:method}

This section presents the model used in the analysis. Subsection \ref{sec:method:model} introduces the ordinal logistic model with category-specific regression parameters, while Subsection \ref{subsec:semi} describes the semi-parallel ordinal logistic model, in which some parameters are shared across categories and others are category-specific.
\subsection{The cumulative ordinal logistic regression model}
\label{sec:method:model}

We modeled the response
$y$ to Question 1 using a categorical variable. 
This variable takes a value of $-1$, $0$, or $1$, corresponding to a perceived negative, neutral, or positive influence of the environment on health. 
This information is accompanied by the individual vector of $P$ covariates $\boldsymbol{x}_i \in \mathbb{R}^P$ related to respondent $i \in \{1, \dots, n\}$.
We treat the response variable as being a realization of a multinomial regression model such that
\begin{align*}
    {y}_i \sim \text{Multinomial}_3(1, {\boldsymbol{\pi}}_i(\boldsymbol{ \eta}_i)),
\end{align*}
 with 
\[
{\boldsymbol{\pi}_i}(\boldsymbol{ \eta}_i) = (\pi_{-1, i}(\boldsymbol{ \eta}_i), \pi_{0, i}(\boldsymbol{ \eta}_i), \pi_{+1, i}(\boldsymbol{ \eta}_i))^\top,
\]
such that $\sum_{j = -1}^{+1}\pi_{j, i}(\boldsymbol{ \eta}_i) = 1$, for any linear predictor $\boldsymbol{ \eta}_i =( \eta_{-1, i},\eta_{0, i})\in \mathbb{R}^2$ having  ``+1'' as reference category. 

One common assumption for dealing with an ordinal response is to equate the cumulative odds to the linear predictor $\boldsymbol{\eta}_i$ as
\begin{align}
\label{eq:cumulativelogits}
  \log\bigg(\frac{\Pr(y_i \leq j)}{\Pr(y_i > j)}\bigg) = \eta_{j, i} =  c_{j} +  \boldsymbol{x}_i^\top \boldsymbol{\beta}_{j},
\end{align}
for $j \in \{ -1,0\}$ and $\boldsymbol{\beta}_{j} \in \mathbb{R}^P$.
 The multinomial probabilities can be easily found from Equation \eqref{eq:cumulativelogits}.

The parameters $c_{j}$ are \emph{ordered intercepts}, such that $c_{-1}<c_{0}$, and are generally referred to as \emph{thresholds} (or cutoffs) of a latent construction of the model \citep{agresti2010analysis}. 
The parameters $\boldsymbol{\beta}_{j}$ are regression parameters, with generic element ${\beta}_{j,p}$ with $p\in \{1, \dots, P\}$. Notably, the regression coefficients $\boldsymbol{\beta}_{j}$ vary across outcome thresholds $j$. Consequently, the overall parameter $\boldsymbol{B}=( {\boldsymbol{\beta}_{-1}} , {\boldsymbol{\beta}_{0}})$ forms a $P\times 2$ matrix. 
Without additional constraints, the parameters in $\boldsymbol{B}$ remain unrestricted.
In this formulation, the model can be viewed as a reparameterization of a multinomial logistic model with constrained intercepts, thereby preserving the natural ordering of the response categories when $\boldsymbol{x} = \boldsymbol{0}$. 
Under this construction, a positive (negative) value of ${\beta}_{j, p}$ indicates a linear increase (decrease) of the cumulative log-odds presented in Equation \eqref{eq:cumulativelogits}, when the $p$--th covariate increases from $x_p$ to $x_p+\delta$, for some $\delta>0$.
As usual in logistic models, this
translates into an increased (decreased) probability of observing $y \leq j$. 

Notably, we are not requiring the typical assumptions of parallelism, also known as the assumption of proportional odds, for which $\beta_{-1,p} = \beta_{0,p}$,  $\forall p \in \{1,\ldots, P\}$.
Since $\beta_{j,p}$ depends on the margin $j$, interpreting a positive value of $\beta_{j, p}$ as a worsening of perception due to an increase in the covariate $x_p$ is not correct or straightforward. Such interpretation also depends on $\beta_{j^\prime, p}$ for $j \neq j^\prime$. 
To clarify the interpretation of the regression parameters, we first note that the parameters $\beta_{j,p}$ and $\beta_{j^\prime, p}$ with $j\neq j^\prime$ can differ in both sign and magnitude. 
Table \ref{tab:coefficients} summarizes the scenarios arising from the possible sign combinations of the parameters.

\begin{table}[]
    \centering
    \begin{tabular}{|c|c|p{4cm}|p{3.5cm}|} \toprule
         $\text{sign}(\beta_{-1, p})$ & $\text{sign}(\beta_{0, p})$ 
         & {Interpretation of  positive change $\delta$ in covariate $(x_p \rightarrow x_p+\delta)$} 
         & {Sketch of why} \\ 
         \midrule
         $+$ & $+$ & 
        The increase in the covariate increases the probability of perceiving the environment as harmful, or either harmful or irrelevant to health.
         &  For both $j \in \{-1, 0\}$, 
            $\Pr(y_i \leq j)$ increases (thus
             $\Pr(y_i> j)$ decreases).
         \\ \midrule
         $-$ & $+$ & The increase in the covariate increases the probability of perceiving the environment as irrelevant to health. 
         
         & $\Pr(y_i \leq  -1 )$ decreases and  $\Pr(y_i \leq  0 )$ {increases};  
         
         hence,  $\Pr(y_i = 0) =
         \Pr(y_i \leq  0 ) - \Pr(y_i \leq  -1 )
         $ increases.
         \\  \midrule
         $+$ & $-$ &
         An increase in the covariate increases the probability of polarized environmental perception, i.e., viewing the environment as harmful or beneficial to health.
         & 
         $\Pr(y_i \leq  -1 )$ { increases} and  $\Pr(y_i \leq  0 )$ { decreases};  hence,  $\Pr(y_i = 0) =
         \Pr(y_i \leq  0 ) - \Pr(y_i \leq  -1 )
         $ decreases.
         \\  \midrule
         $-$ & $-$ & The increase in the covariate increases the probability of perceiving the environment as beneficial or not harmful to health.
         & For both $j \in \{-1, 0\}$, 
            $\Pr(y_i \leq j)$ decreases (thus 
             $\Pr(y_i>  j)$ increases).
             \\ \bottomrule
    \end{tabular}
    \caption{Interpretations of the signs of estimated coefficients with respect to positive changes in the covariate $x_p$, fixing the others.}
    \label{tab:coefficients}
\end{table}

In brief, the parameter $\boldsymbol{\beta}_p = ({\beta}_{-1,p}, {\beta}_{0,p})^\top$ of each covariate can be classified according to the Cartesian quadrant in which it lies.
The first quadrant ($\beta_{-1,p} > 0$,  $\beta_{0,p} > 0$) indicates increased probabilities of both negative and neutral environmental quality perception as $x_p$ increases.
The second quadrant ($\beta_{-1,p} < 0$,  $\beta_{0,p} > 0$) is associated with an increased tendency toward a neutral environmental quality perception when $x_p$ increases.
The third quadrant ($\beta_{-1,p} < 0$, $\beta_{0,p} < 0$) indicates a decreased probability of both negative and neutral perceptions, and hence a shift toward a positive feeling.
Finally, the fourth quadrant ($\beta_{-1,p} > 0$,  $\beta_{0,p} < 0$) indicates a reduced tendency toward neutrality, with responses polarized between negative and positive perception.
This may reflect ambivalence or a lack of clear positioning that
leads people with similar characteristics to respond in opposite ways.


Notably, coefficients that lie close to the bisector between the first ($\beta_{-1,p} > 0$, $\beta_{0,p} > 0$)  and third ($\beta_{-1,p} < 0$, $\beta_{0,p} < 0$) quadrants indicate unidirectional changes toward positive or negative perception.
The limiting case under which all coefficients lie in this bisector line is the \emph{parallel odds} model.
This specification assumes that changes in $\boldsymbol{x}$ produce equal shifts in both log-odds functions.
Under these restrictions, the link between the covariates and the cumulative odds is represented by the linear predictor
$    \eta_{j, i} =  c_{j}+  \boldsymbol{x}_i^\top \boldsymbol{\beta},
$ 
where only the intercepts (or thresholds) vary across outcome margins, while the regression coefficients remain constant.
Despite its widespread use in practice, the simplicity of this model comes at the cost of restrictions that are often too stringent, especially given the complexity of survey data and the variety of scenarios illustrated in Table \ref{tab:coefficients}.

While test procedures for the \emph{parallel odds} assumption are typically conducted using the Brant test \citep{BrantTest}, the context often indicates which covariates are subject to the parallelism assumption.
In such cases, a commonly applied approach is the \emph{partially parallel odds} model \citep{fullerton_xu_ordered_2016}, in which certain parameters are constrained to be equal across categories, while others are allowed to vary freely.
In the absence of this substantial knowledge, it is possible to adopt the \emph{semi-parallel odds} model proposed by \cite{JSSordinalElNet}, based on a penalized likelihood approach, as explained in the next subsection.

\subsection{The semi-parallel cumulative ordinal logistic regression model}\label{subsec:semi}

Although the non-parallel cumulative ordinal regression model, presented in Section \ref{sec:method:model}, offers several interpretative advantages previously discussed, it is important to recognize its limitations.
First, the inclusion of multiple covariates can lead to computational issues such as separation or quasi-separation, resulting in empty cells or outcome categories with only a single response modality.
These issues are particularly common when multiple dichotomous covariates and their interactions are included.
Such problems are well documented in logistic regression \citep[see, e.g.][among others]{logisticRidge1, ridgelogistic}, and are commonly addressed through penalization methods that shrink parameter estimates toward zero \citep{harrell2015regression, L1ordinal}.
Second, the evident correlations among our predictors raise natural concerns about multicollinearity and then reliability about uncertainty measures. Standard remedies such as dimensionality reduction techniques often sacrifice interpretability, while pre-selecting a limited set of predictors is not feasible, as in our application, where multiple correlated covariates are expected to jointly explain variability in the response. 
Penalization, therefore, provides a natural and coherent strategy to address both separation and multicollinearity in this context \citep{harrell2015regression, L1ordinal}.

We therefore employ the semi-parallel model proposed by \cite{JSSordinalElNet}, an elastic-net–penalized formulation that provides a compromise between parallel and non-parallel ordinal models, while addressing multicollinearity and high-dimensional predictor spaces through regularization.
Let define 
\begin{align}
\label{eq:double_pred}
  \eta_{j,  i}= c_{j} +  \boldsymbol{x}_i^\top \boldsymbol{\beta}_{j},
\end{align}
and decompose $\boldsymbol{\beta}_{j}$ as $\boldsymbol{\gamma} +\boldsymbol{\gamma}_{j}$. We then account for both shared $\boldsymbol{\gamma}$ and margin-specific effects $\boldsymbol{\gamma}_{j}$, stored in $ \boldsymbol{\Gamma} =\begin{bmatrix}
  \boldsymbol{\gamma}_{-1} & \boldsymbol{\gamma}_{0}
\end{bmatrix}$. 

The objective function to be minimized 
\begin{align*}
    \mathcal{M}(\boldsymbol{c},\boldsymbol{\gamma}, \boldsymbol{\Gamma}) = - \mathcal{L}( \boldsymbol{c},\boldsymbol{\gamma}, \boldsymbol{\Gamma}; \boldsymbol{y}, \boldsymbol{X}) +  \mathcal{P}(\boldsymbol{c},\boldsymbol{\gamma}, \boldsymbol{\Gamma}; \lambda, \alpha,\rho),
\end{align*}
 is a sum of a multinomial log-likelihood  $\mathcal{L}$ and a penalty function $\mathcal{P}$.
We denote with $\hat{\boldsymbol{\boldsymbol{c}}}$, $\hat{\boldsymbol{\gamma}}$, $\hat{\boldsymbol{\Gamma}}$, the results to the  minimization problem, and set $\hat{\boldsymbol{\beta}_j} = \hat{\boldsymbol{\gamma}}+ \hat{\boldsymbol{\gamma}}_j$.
The weighted and rescaled multinomial log-likelihood equals
\begin{align*}
    \mathcal{L}( \boldsymbol{c},\boldsymbol{\gamma}, \boldsymbol{\Gamma} ; \boldsymbol{y}, \boldsymbol{X}) = \frac{1}{n}\sum_{i=1}^{n}  w_{i}\Big[ \sum_{j=-1}^1  \mathbbm{1}(y_i = j) \log\big(\pi_{j}(\boldsymbol{\eta}_i)\big)\Big],
\end{align*}
with sampling weights $w_{i}$ that satisfy $\sum_{i=1}^n w_i = n$, and $\pi_{j,i}(\boldsymbol{\eta}_i)$ and the linear predictor $\boldsymbol{ \eta}_i$ as defined in Equation \eqref{eq:double_pred}.

The penalty function $\mathcal{P}(\boldsymbol{c},\boldsymbol{\gamma}, \boldsymbol{\Gamma}, \lambda, \alpha,\rho)$
is defined as
\begin{align*}
   \mathcal{P}(\boldsymbol{c},\boldsymbol{\gamma}, \boldsymbol{\Gamma}, \lambda, \alpha,\rho) = \lambda \bigg( \rho \sum_{p=1}^{P} \big(\alpha |\gamma_{p}| + \frac{1}{2} (1-\alpha)\gamma_{p}^2 \big)  + \sum_{p=1}^{P} \sum_{j=-1}^0 \big( \alpha |\gamma_{j, p}| + \frac{1}{2}(1-\alpha) {\gamma_{j, p}^{2}}\big)\bigg),
\end{align*}
where $ \lambda \geq 0$, $\alpha \in [0,1]$, $\rho \geq 0$ play the role of tuning parameters, controlling the degree of regularization toward three possible directions. 
The tuning parameter $\lambda$ controls the overall degree of penalization \citep[see, e.g.][]{tibshirani_lasso_1996}. 
When $\lambda =0$, the estimation relies solely on the likelihood without any penalization. As $\lambda$ increases, the elastic-net penalty contributes more heavily to the estimation, by shrinking both $\boldsymbol{\gamma}$ and $\boldsymbol{\Gamma}$ to zero.
Although understanding the role of $\lambda=0$ is somewhat didactic, setting it to such a value is not recommended, since the model then suffers from non-identifiability issues due to the double additive structure of the linear predictor as defined in Equation \ref{eq:double_pred}.
Instead, the tuning parameter $\alpha$ controls the balance between the \emph{lasso} ($\alpha = 1$) and the \emph{ridge} ($\alpha = 0$) penalties. Large values favor models with many parameters exactly shrunk to zero, with the others left to vary. 
This is a nice feature when the interest is in selecting variables and reporting a model with few, interpretable effects.
Small values lead to models that are intrinsically less parsimonious,
but still can handle multicollinearity, leading to more stable estimates where all variables contribute to some extent.
Finally, the tuning parameter $\rho$ has the role of calibrating the results between two extremes.
{ According to \cite{JSSordinalElNet}, $\rho = 0$
 will leave the parallel coefficients unpenalized, so the fit will shrink from the maximum likelihood nonparallel model to the fit toward the maximum likelihood parallel model fit as $\lambda$ increases from zero.
 Vice versa, when $\rho$ is fixed at a very large value, higher penalization is given to parallel coefficients, so that the model becomes equivalent to the nonparallel model.
 
 For further technical details of the penalization and maximization procedure, we refer to  \cite{JSSordinalElNet} original work.}
The choice of hyperparameters  can be guided by cross-validation and goodness-of-fit criteria, while variability can be estimated by bootstrapping. As these aspects are application-related, we defer their discussion to Section \ref{sec:res}.

\section{Application}\label{sec:res}

This section introduces the model specification (Subsection \ref{subsec:spec}), followed by the related results (Subsection \ref{subsec:res}), and concludes with a brief discussion on policy implications and approach limitations (Subsection \ref{subsec:policy}).

\subsection{Model specification and uncertainty quantification}
\label{subsec:spec}
\subsubsection{Model specification}
Here we define the linear predictor in Equation~\eqref{eq:double_pred} as a sum of two sources of information: 
(i) a linear functional form of the national-level  effects, where the covariate vector $\boldsymbol{m}_i$  captures both first-order marginal effects of covariates defined in Table \ref{tab:variables} (excluding LHU membership), as well as all their two-way interactions;
(ii) the subject’s membership in one of $L$ LHUs, represented by the vector $\boldsymbol{l}_i = \begin{bmatrix}
    \mathbbm{I}(l_i=1) &  \ldots & \mathbbm{I}(l_i=L) 
\end{bmatrix}^\top,$ which determine the local effect. 
More specifically, let $\boldsymbol{\beta}_j$ be the regression margin-specific parameter of interest.
We divide $\boldsymbol{\beta}_j$ and $\boldsymbol{x}_i$ in two parts, i.e., $\boldsymbol{\beta}_j = \begin{bmatrix}
      \boldsymbol{\beta}_{\text{M},j}^\top & \boldsymbol{\beta}_{\text{L},j}^\top
\end{bmatrix}^\top$ and $\boldsymbol{x}_i = \begin{bmatrix}
      \boldsymbol{m}_i^\top & \boldsymbol{l}_i^\top
\end{bmatrix}^\top$.
The linear predictor for margin $j$ is written as 
\begin{align*}
    \eta_{j,i} = c_j + \boldsymbol{\beta}_{\text{M},j}^\top \boldsymbol{m}_i + \boldsymbol{\beta}_{\text{L},j}^\top  \boldsymbol{l}_i.
\end{align*}

The linear combination $\boldsymbol{\beta}_{\text{M},j}^\top \boldsymbol{m}_i$ defines the \emph{national-effects} component of the model, which captures variability in $y_i$ through $\boldsymbol{m}_i$. The vector $\boldsymbol{m}_i$ comprises both individual-level, municipality-level covariates and the interactions. 
The coefficients $\boldsymbol{\beta}_{\text{M},j}$ are assumed to be homogeneous across the country, capturing the main sources of between-individual variability. 
Such variability arises both from individual characteristics and contextual factors (captured by information on municipalities). 
Including second-order interactions further accounts for heterogeneity in contextual risk effects. 
In particular, variation in municipality-level exposures (e.g., pollution, type of municipality) may yield differential impacts conditional on individual-level attributes (e.g., education, economic status), holding other factors fixed. 
These interaction terms quantify subgroup-specific sensitivities to environmental risks and capture systematic effect modification across the population, which may reflect differences in vulnerability or awareness.

The linear combination $\boldsymbol{\beta}_{\text{L},j}^\top  \boldsymbol{l}_i$ 
is a fixed effect at the LHU level used to capture systematic deviations from the national effects component for margin $j$.
 We choose the LHUs as the finer aggregation level since they are the geographical units of the sample design \citep[][]{iss_2023_protocollo_passi}.
 LHUs are a compromise between capturing variability at the micro-territorial scale (i.e., municipalities) and at broader macro-territorial levels (e.g., regions). 
In Italy, LHUs are public entities with organizational, managerial, and technical autonomy, responsible for delivering healthcare services within a specific sub-regional territory.
Since Question~\ref{q:\thequestion} is formulated in general terms, identifying these deviations may point to contextual factors that are not explicitly included in the model, but influence perception consistently across the entire LHU territory. 
The observed deviations may indicate localized issues to investigate, influencing well-being and healthy living in the surrounding environment. 

\subsubsection{Choice of hyperparameters and uncertainty quantification}

Some aspects remain unaddressed in the model proposed above, i.e., the selection of the hyperparameters. 
As discussed in Subsection \ref{subsec:semi}, the hyperparameters $\lambda$, $\alpha$, and $\rho$ regulate, respectively, the overall degree of penalization, the ridge–lasso trade-off, and the parallel–nonparallel assumption.
While setting $\alpha = 0.5$ provides a compromise between non- and too conservative covariate selection, the practical choice of $\lambda$ and $\rho$ is not straightforward.
Hyperparameters were selected using $20\%$ of the data for training, which was excluded from subsequent analyses.
We conducted a grid search over values of $\log
(\lambda) \in \{-13.82, -12.28, -10.75,  -9.21,  -7.68, -6.14,  -4.61\}$ and $\rho \in \{0.05, 0.46, 0.87, 1.28, 1.68, 2.09, 2.50\}$, and compared the out-of-sample performance using five-fold cross-validation and the Ranked Probability Score (RPS) metrics.
The RPS is a well-known metric for evaluating the probabilistic forecasts of ordinal models, which assesses the entire shape of the predicted distribution.
In total, we investigated $49$ configurations with varying degrees of penalization.
Out-of-sample performance metrics are reported in the SM.

In the application, we selected $\rho = 1.275$ and $\log(\lambda) =  -7.68$.
This choice does not correspond to the optimal value in the analyses (which, based on the average RPS, would be $\log(\lambda) =  -7.68$ and $\rho =2.5$), but represents a compromise to avoid substantial losses in the case of excessive penalization.
Indeed, a large $\rho$ leads to considerable losses when $\lambda$ becomes large, while it appears to have little influence when $\lambda$ decreases. 
We then set $\lambda$ to be the largest value that does not produce substantial drops, selecting an intermediate $\rho$ 
to prevent us from large losses associated with substantial drops in predictive performance when $\lambda$ is set large.
Additional details are reported in the SM, along with a graphical assessment of how the effects of interest change under alternative hyperparameter specifications.

To quantify the variability of parameter estimates and performance measures, we implement a bootstrap procedure that respects the stratified nature of the sample \citep{rao_woo_bootstrap_1988}.
{ The bootstrap procedure is performed by estimating the model on $80\%$ of the data not used for hyperparameter selection. 
} For each replication $r = 1, \ldots, R$, and for each stratum $s = 1,\ldots,S$ with $n_s$ respondents--defined by the combination of LHU, sex, and age class--we resample $n_s$ units with replacement to obtain the set of units $\mathcal{D}_{r,s}$ with $|\mathcal{D}_{r,s}| = n_s$. 
The full bootstrap sample set at replication $r$ is then given by $\mathcal{D}_r = \{\mathcal{D}_{r,1},\ldots,\mathcal{D}_{r,S}\}$. 
Under the chosen hyperparameters, the model is then estimated with the \texttt{ordinalNet} \texttt{R} package \citep{JSSordinalElNet}, and the regression parameters $\hat{\boldsymbol{c}}_r$ and $\hat{\boldsymbol{B}}_r$ are stored.  All graphs and derived metrics are obtained from these replications. { Pseudocode of bootstrap procedure is reported in the SM}.

{
We also performed a predictive evaluation using 5-fold cross-validation with RPS and Misclassification Error (ME), to assess the model’s performance against several competing approaches. All the results are reported in the SM.

}

\subsection{Main results}
\label{subsec:res}

\subsubsection{Explanatory performance of the model}
{
We introduce here the model by evaluating its explanatory ability, highlighting both its potential and limitations in producing meaningful insights.}
We start by defining the { variance partition coefficient (VPC)} for margin $j$, as an approximating measure the explained variance by the model over the total \citep{goldstein2002partitioning,browne2005variance}.
This can be derived by considering the well-known latent-variable representation of logistic regression \citep{agresti2010analysis}.

For margin $j$, we computed the explained variance as
\begin{align*}
\hat{\sigma}_{j}^2 =  \frac{1}{n-1}  \sum_{i=1}^{n}(  \hat{\eta}_{j, i}-\bar{\eta}_j)^2,
\end{align*}
where $\bar{\eta}_j$ denotes the mean for margin $j$. 
This variance then enters into the $\text{VPC}_j$ formulation as:
\begin{align*}
   \text{VPC}_j= \frac{\hat{\sigma}_{j}^2 }{\hat{\sigma}_{j}^2 + \pi^2/3},
\end{align*}
where $\pi^2/3$ is the variance of a standard logistic distribution. Similar formulas can be derived by considering specific components of the model (e.g., LHU, national effects, LHU and national effect covariance).
We report the estimated value of the $\text{VPC}_j$ in Table \ref{tab:variance}, together with the quantification of the different variance components involved in the model defined in Subsection \ref{subsec:spec}.

\begin{table}[]
    \centering
    \begin{tabular}{|l|ccccc|c|}
    \hline 
         {Margin} $j$ & $\hat{\sigma}_{j}^2 + \pi^2/3$  & $\hat{\sigma}_{j}^2$ & $\hat{\sigma}_{j,\text{L}}^2$ & $\hat{\sigma}_{j,\text{M}}^2$ & $100 \times \text{VPC}_j$  \\ 
         \hline         $\Pr(y\leq-1)$
         &  3.92 (0.04) & 0.63 (0.04) & 0.34 (0.03) & 0.24 (0.03) & 16.06 (0.87)  \\
                  $\Pr(y \leq 0 )$ 
                  & 4.38 (0.06) & 1.09 (0.06) & 0.96 (0.06) & 0.14 (0.02) & 24.97 (1.13)\\
                  \hline
    \end{tabular}        
    \caption{Contribution to the total variance of each model component, in its latent variable representation. 
    The values in parentheses represent the standard deviations, computed via bootstrap with $R=5500$ replications.}
    \label{tab:variance}
\end{table}
We note that the model explains approximately $16\%$ of the variance for the linear predictor modeling strictly negative responses, and about $25\%$ for the predictor accounting for both neutral and negative responses. 
{
The fact that a considerable amount of variance remains unexplained highlights the inherent difficulty in explaining environmental perception. 
Such difficulty is also confirmed by the informational gain that we obtain in the cross-validation reported in the SM.
The penalized models are the best in terms of predictive power, but the most substantial gains emerge when spatial location information (e.g., LHUs belonging) is included in the models.
The advantage of including individual information, or individual and contextual information together, emerges in small out-of-sample predictive gains, highlighting how, overall, territoriality associated with LHUs is the main source of variability.
}

{ Beyond predictive capabilities, the proposed model allows for a detailed exploration of the underlying determinants of variability. We}
 note how $88\%$ of the explained variance is attributable to the LHUs' fixed-effects for $\Pr(y \leq 0)$, and $53\%$ $\Pr(y=-1)$. 
Therefore, for $\Pr(y \leq 0)$, one may conclude that the individual and contextual covariates used have limited explanatory power. 
At the same time, a strong territorial/local component captured by the fixed effects of LHU is present.
By contrast, in the case of $\Pr(y=-1)$, the covariates used can explain a non-negligible part of the variability, highlighting their effective relationship with a strictly negative perception.
This disparity between responses is interesting and potentially explainable by multiple concurrent factors linked to the nature of the responses and the selected covariates.
A negative response is a clear and unambiguous indicator of recognition of a potential problem for health. 
In contrast, modelling $\Pr(y \leq 0)$, which can also be expressed as $1-\Pr(y =1)$, introduces greater ambiguity. Neutrality ($y=0$) may be interpreted as the absence of problems and thus, in a positive light, similar to $y=1$.
Moreover, some contextual covariates selected for model construction represent actual concerns when their levels are high. For example, higher pollution levels are associated with strong negative impacts on health. However, the absence of these factors does not necessarily imply positive perceptions. 

\subsubsection{LHUs fixed effects}
\begin{figure}[t]
  \centering
  \includegraphics[width=1\textwidth]{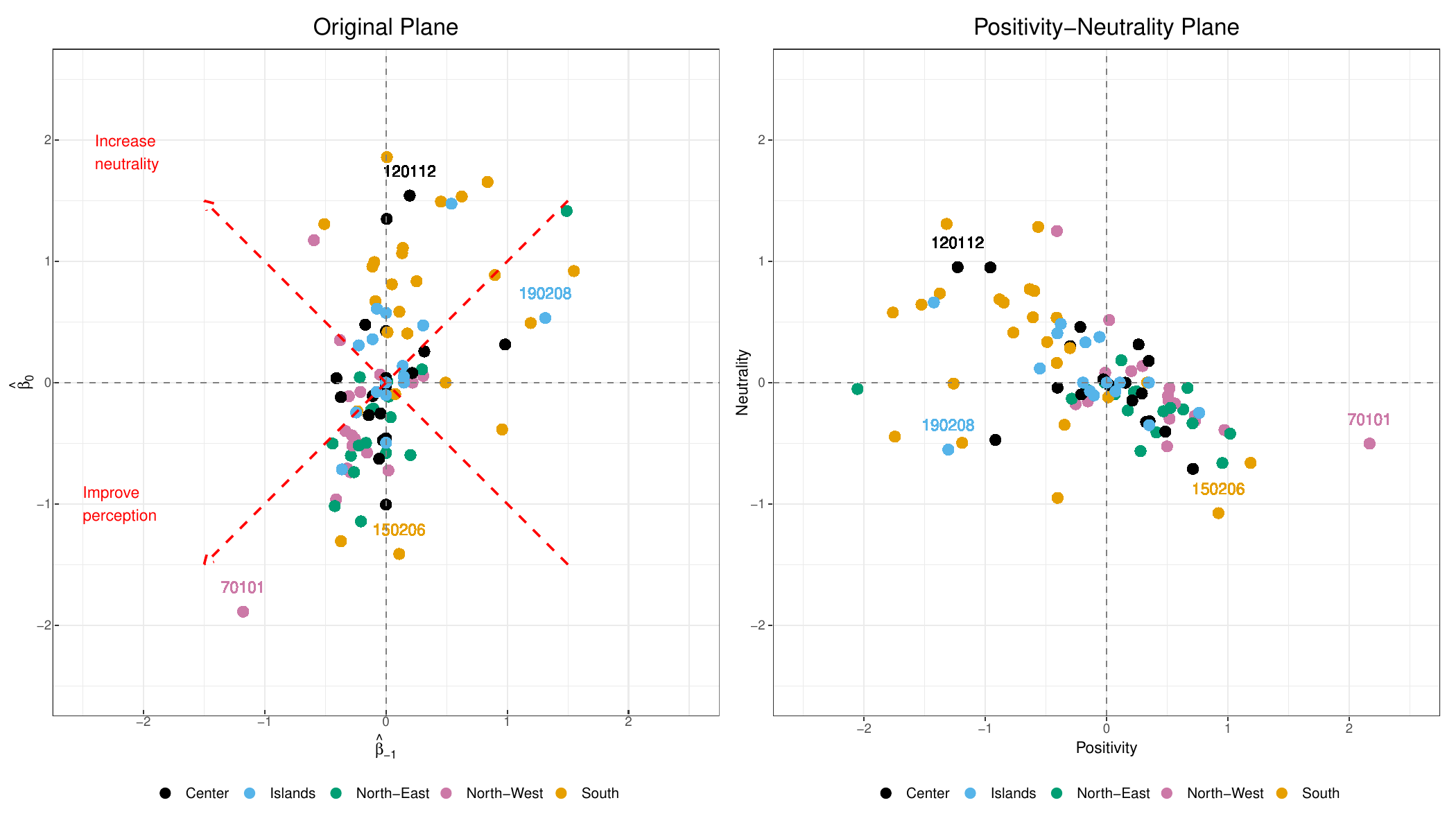}
  \hfill
  \caption{On the left, point estimates of coefficients $\hat{\boldsymbol{B}}_\text{L} = \begin{bmatrix}
\hat{\boldsymbol{\beta}}_{\text{L},-1} & \hat{\boldsymbol{\beta}}_{\text{L},0}  \end{bmatrix}$ regarding different LHU fixed effects. On the right, the coefficients are rotated by $135^\circ$ and flipped in their second dimension. Colors identify different macro-regions in Italy. Text highlights selected LHUs.}
  \label{fig:plane_analysis2}
\end{figure}

We now focus on outputs of the model related to the LHUs' fixed effects. 
In Figure~\ref{fig:plane_analysis2}, the left panel displays the estimated coefficients for each LHU from the selected semi-parallel ordinal model. 
The $x$-axis represents the estimated effect on the log-odds scale of responding ``negative'', i.e., $\text{logit}(y_i = -1)$, while the $y$-axis represents the effect on the log-odds of responding ``negative or neutral'', i.e., $\text{logit}(y_i \leq 0)$.

As each point corresponds to an LHU, its position in the plane also reflects the estimated direction and strength of the association with the perception categories, as summarized in Table \ref{tab:coefficients} and explained in Section \ref{sec:method}.
Under the parallel assumption, all coefficients would lie along the bisector.
Interestingly, the estimates of the fixed LHUs coefficients show substantial vertical spread, indicating departures from this assumption.
This departure highlights a greater territorial diversity in distinguishing the positive response from the other response modalities. 
In contrast, we note that the coefficients are not widely dispersed along the horizontal axis, suggesting limited territorial heterogeneity in explaining the negative response relative to the other modalities. 
We note that only a few estimated LHUs coefficients deviate more markedly.
 
The right panel shows the same information, but with estimated coefficients that were rotated by $135^\circ$ and sign-flipped in the second dimension.
This transformation positions the estimated coefficients in the \emph{positivity-neutrality} plane.
A parallel model places all covariate effects on a single latent axis, here termed the \emph{positivity} dimension. 
For any covariate, the sign and magnitude of its coefficient determine its position on this dimension: a positive coefficient implies that increases in the covariate are associated with higher probabilities of more positive perception outcomes, whereas a negative coefficient indicates an overall shift toward less positive outcomes.
Departures from the parallel assumption are captured by loadings on an axis orthogonal to the positivity one, which we call the \emph{neutrality} dimension.
The neutrality dimension reflects non-parallel, cut-point–specific effects, indicating whether a covariate mainly increases neutrality (centralization) or reduces it (polarization) rather than producing a uniform shift from negative to positive categories.

In these terms, we recognize LHUs with estimated effects associated with 
increased positivity, while altering the baseline neutrality with different degrees (e.g., 70101, Imperiese, Liguria or  150206, Napoli 3 Sud, Campania);
LHUs with increased neutrality but decreased positivity  (e.g., 120112, Frosinone, Lazio); 
and those with both decreased positivity and neutrality (e.g., 190208, Siracusa, Sicilia).
Since this latter group of LHUs is characterized by negative values in both positivity and neutrality, we place particular emphasis on them, as respondents expressed an unexpectedly negative perception, potentially highlighting environmental issues.

\begin{figure}
    \centering
    \includegraphics[width=0.485\linewidth]{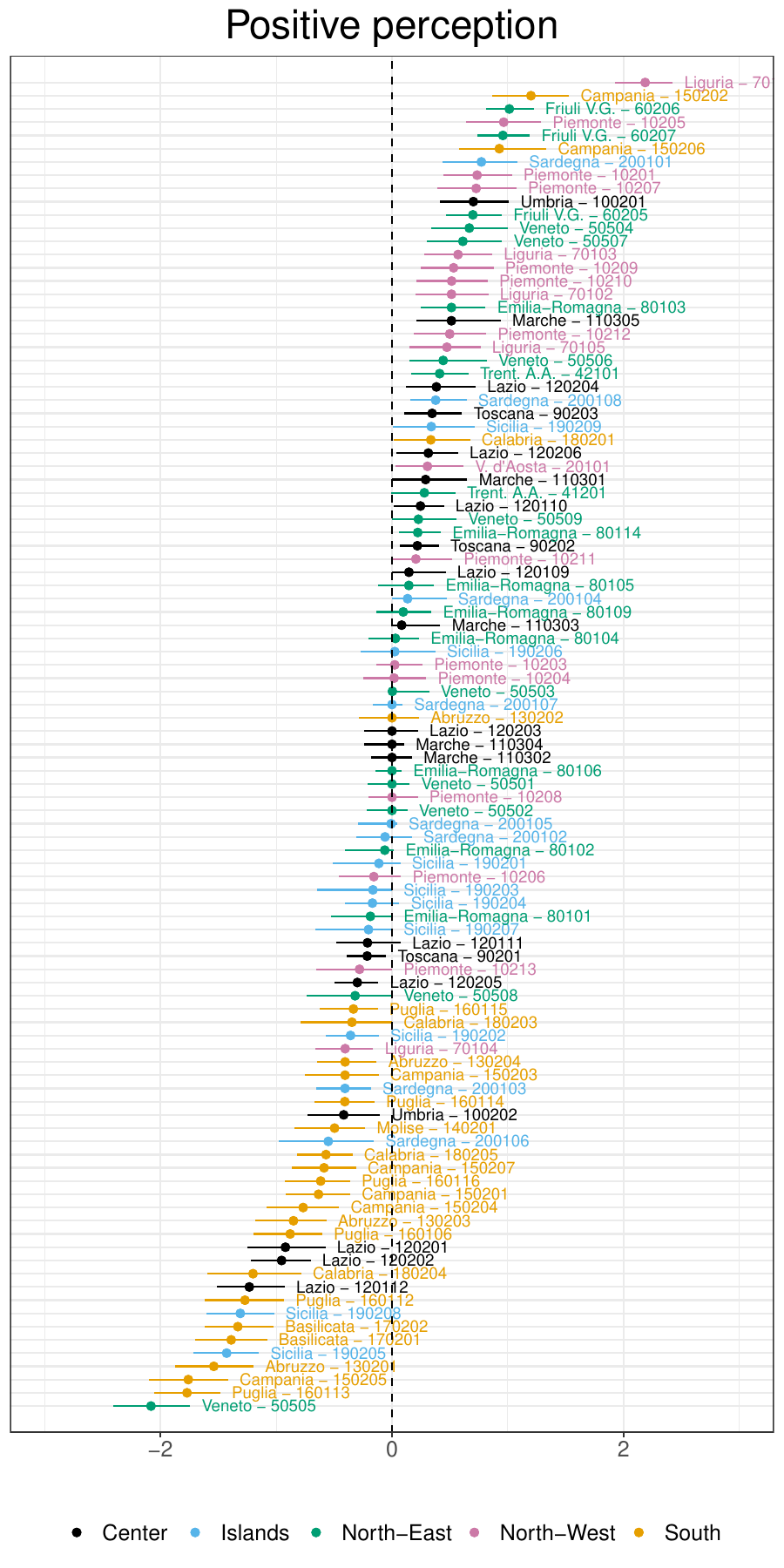}
 \includegraphics[width=0.485\linewidth]{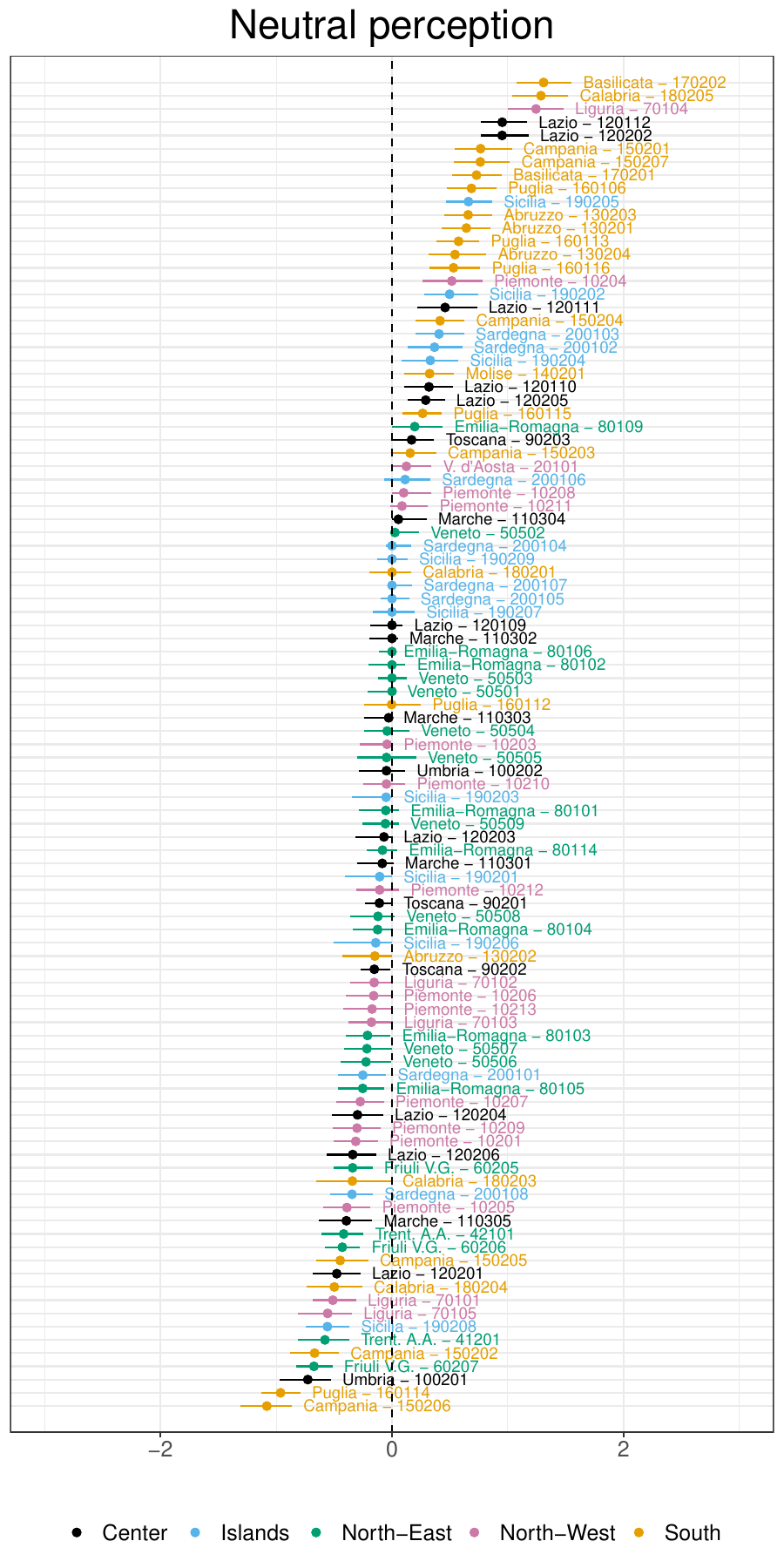}   \caption{Ranking of LHUs based on the positivity and neutrality coordinates, i.e., after rotating by $135^\circ$ and flipping them in their second dimension. Colors identify different macro-regions in Italy.   
Bands report $95\%$ marginal confidence intervals.}
    \label{fig:asl_effects}
\end{figure}

In addition, Figure \ref{fig:asl_effects} reports the LHUs coefficients in the two dimensions of the {positivity-neutrality} plane, with the respective $95\%$ marginal confidence bands obtained via bootstrap with $R=5500$ replications. 
The left panel displays the results for the positivity dimension, while the right panel shows those for the neutrality one.
{
These transformed estimated effects are placed in descending order with respect to their pointwise values in the two derived dimensions.
We highlight that the ranking obtained should not be interpreted as a direct measure of which LHUs exhibit better or more neutral environmental perception, but rather as deviations from the constructed national-effects model.
Further, no corrections for multiple testing have been applied.
Although this issue is methodologically and practically relevant for inference, a thorough investigation is deferred to future work that can fully address the complexity of the penalized framework \citep[see, e.g.,][]{meinshausen2009p, wasserman2009high}, highlighting the exploratory nature of the present study.
}

By looking at the figures, we note how a dependence on the macro-areas of the territory appears evident, especially when comparing Northern LHUs (colored in pink and green) with the Southern (colored in yellow), with some exceptions. 
A North-South gradient is often observed in studies concerning Italy in various fields of research,
including health \citep[e.g.][]{petrelli2024socioeconomic,stival2024bayesian}, and environmental studies \citep[e.g.][]{ISTAT_env, pasetto2022environmental}. This gradient seems to be more pronounced along the positive perception dimension. 
Interestingly, although contextual and individual covariates characterized by macro-territorial heterogeneity were included in the model (cf. Figure~\ref{fig:positivity_determinants}), these were not sufficient to eliminate such disparities, thereby opening the door to future developments.

The rankings represented in Figure \ref{fig:asl_effects} provide a basis for further thoughts.
At the lower end of the ranking, we can identify those LHUs associated with a stronger estimated negative perception.
According to the right panel, we identify LHUs with larger estimated deviations from the baseline model, e.g., LHU 50505 (ASL del Polesine),  LHU 160113 (Barletta-Andria-Trani),  LHU 150205 (corresponding to the Northern Naples area).
In so doing, however, we check whether these deviations are accompanied by an estimated increase in neutrality.

LHU 50505 is characterized by negative coefficients on the positivity dimension and by a neutrality coefficient close to zero, with  moderate estimated variability. 
The  Polesine health unit area is located in the south-eastern part of the Veneto region, along the mouths of the Adige and Po rivers. 
Residents in this area may perceive additional concerns not directly captured in the present analysis (e.g., water pollution, risk of flooding).
Moreover, according to  \cite{istat_2024_veneto_best}, this area was classified in $2023$ as one of the most disadvantaged in the region in terms of several socio-economic indicators. 
ISTAT grouped it with Belluno (LHU 50501, Dolomiti), which, however, represents a special case due to its distinctive geography, being predominantly mountainous.

 LHU 150205 shows negative estimates for both positivity and neutrality, revealing a marked shift toward negative perceptions with respect to the baseline model.
According to \url{https://www.cittametropolitana.na.it/zone-omogenee}, this LHU encompasses two distinct homogeneous administrative areas. 
The first one, which includes the municipality of Quarto, Pozzuoli, and Procida, among others, is the Flegreo-Giuglianese area. 
This areas is characterized primarily by efforts in environmental qualification and requalification. 
These includes  challenges related to the environmental restoration in the Giuglianase area, protection of the territorial basin in the volcanic region of the Phlegraen Fields, as well as initiatives for cultural, environmental, and eco-naturalistic valorization. 
The second one, which includes municipalities of Marano di Napoli, Caivano, Afragola, among others, is one of the most densely populated zones of Southern Italy. 
This area contains the majority of the industry of the Metropolitan City of Naples.
Given the general formulation of the question, its estimated shift towards negative perception may be partially explained by the peculiar nature of these LHUs. So, the covariates and the functional relationships of the linear predictor may not fully capture their complex socio-economic and geographical characteristics.

\subsubsection{National effects}

\begin{figure}
    \centering
    \includegraphics[width=0.475\linewidth]{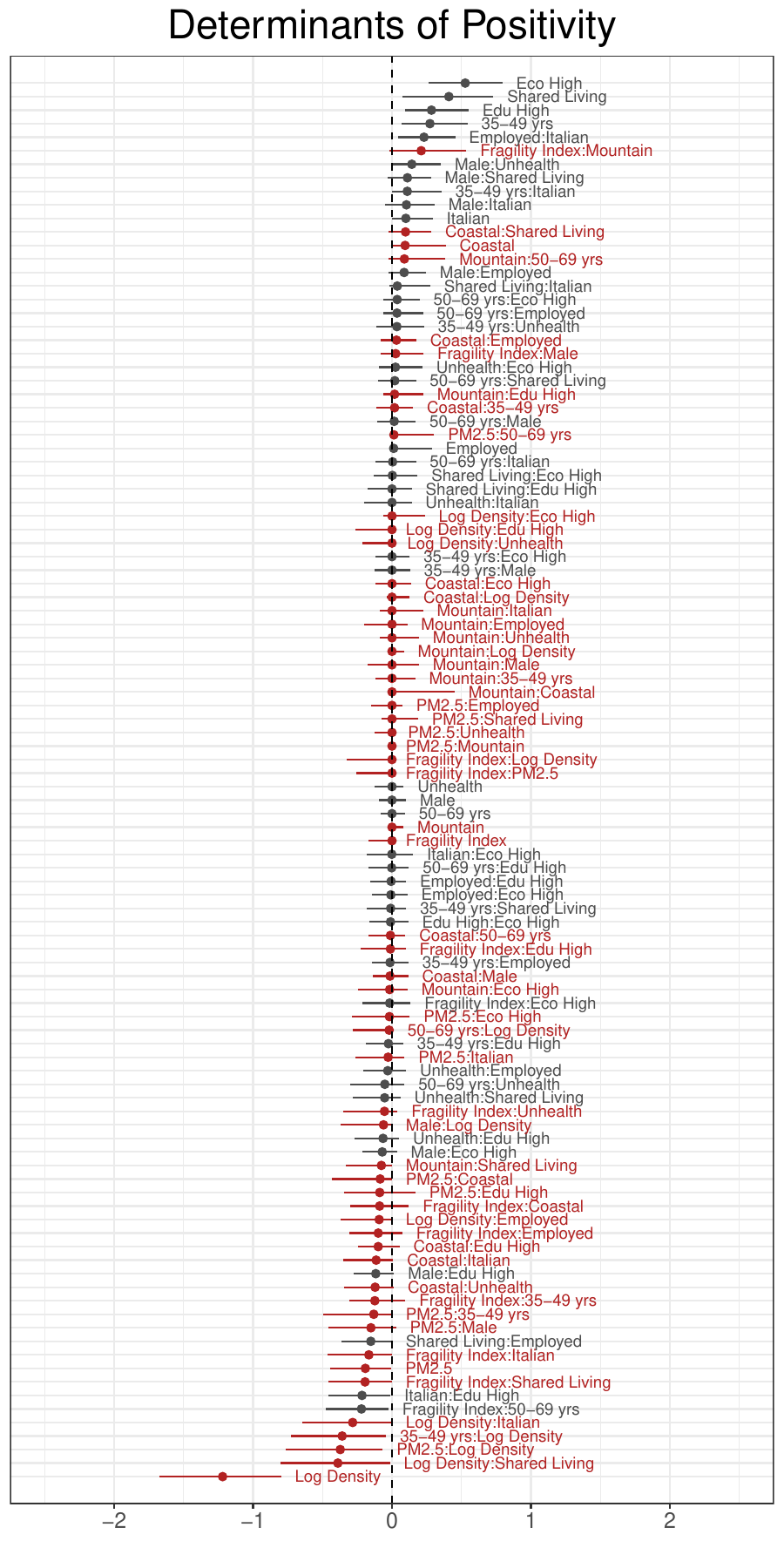}\includegraphics[width=0.475\linewidth]{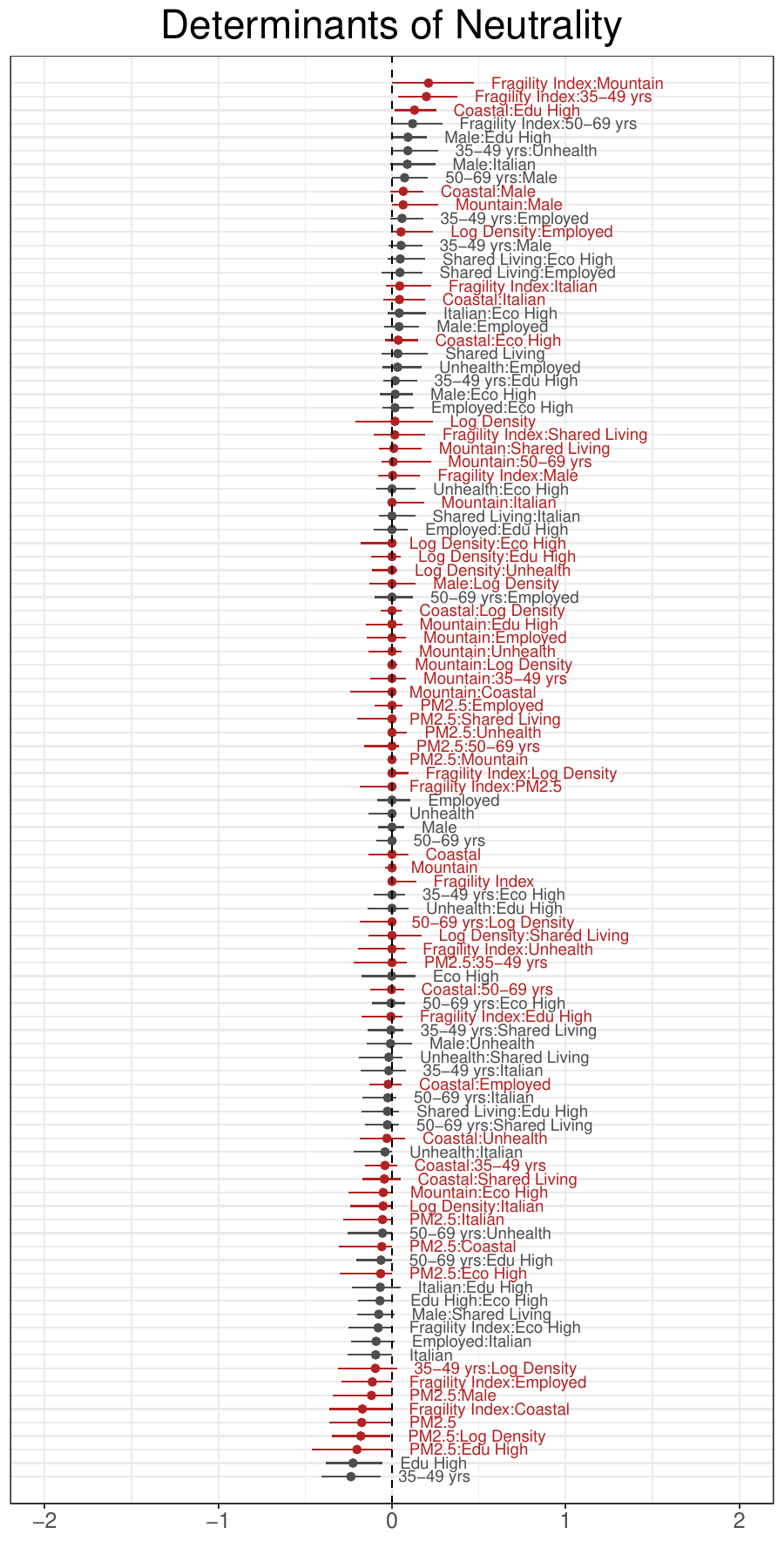}
        
    \caption{The estimated effects $\hat{\Gamma}_\text{M}$ represented in the positivity and neutrality coordinates, i.e., after rotating by $135^\circ$ and sign-flipping them in their second dimension.
    Red color identifies coefficients measuring main effects and interactions with the PM2.5 average pollution level.}
    \label{fig:positivity_determinants}
\end{figure}

{
This subsection examines the coefficients of the main national model, which describes how individuals’ perceptions vary in response to changes in individual- and contextual-level covariates, and their interactions.

Figure \ref{fig:positivity_determinants} reports the rotated coefficients and their corresponding 95\% marginal uncorrected confidence intervals, ranked in descending order by point estimates. 
 Colors are used to distinguish coefficients that refer exclusively to individual covariates from PASSI (grey) from those related to contextual territorial factors (red), including interactions with individual characteristics.}

{
 The figure provides a comprehensive overview of the behavior of estimates in both directions, facilitating the identification of relevant patterns for general guidance. 
We highlight that the figure should be interpreted with caution, as it displays penalized estimates of effects that differ in nature---namely, categorical variables as dummies, continuous variables, and pairwise interactions---all with uncorrected confidence intervals. 
For instance, strictly negative estimated effects occur with higher proportion among those regarding contextual variables, with respect to exclusively individual covariates (bootstrap distribution of group proportions reported in the SM). 
This result suggests the ability of the selected contextual variables in reflecting actual respondents' concerns (negative perception).

}
We note how increased positivity seems to be associated pointwise with individual characteristics such as declaring no economic difficulties in making ends meet, living with relatives or others,  higher education, being aged 35-49. 
While the sign of some of these estimates are expected (e.g., declaring no economic difficulties in making ends meet), we highlight that the presence of interactions with opposite signs partially adjusts the estimated positive effects for some population subgroups. 
Examples include the negative estimated effect among individuals aged 35-49 living in highly populated municipalities or among highly educated Italians.
{
On the contrary,  decreased positivity estimates seem associated with negative contextual risk factors, particularly those linked to high levels of urbanization (log-density and pollution) placed at the bottom of the figure. 
}{ By looking at the right panel, pollution seems also associated with a slight decrease in neutrality.
}According to the point estimates, this decrease in neutrality seems more pronounced among highly educated people, and among highly educated people in highly polluted municipalities.

{
Even if the evidences discussed above appears to be weak, as some of the intervals include, or nearly include, zero, without any multiplicity correction, discussing these findings is still interesting within their broader context and policy implications, while also outlining current limitations and opportunities for future research. 
}

For the sake of brevity, we report in SM alternative plots that show exposure-probability curves in different regions, thereby highlighting the presence of within-region variability.

\subsection{Discussion on policy implications and future directions}
\label{subsec:policy}
{
The data-integration approach combined with penalization appears adequate for characterizing the context in which people live, as the contextual information yields meaningful estimates and, as shown in the SM, leads to improved out-of-sample predictions compared to models that exclude it or rely on alternative strategies.
Despite these gains, it is nevertheless essential to emphasize the challenges encountered and to discuss the results within a broader contextual framework. 
This approach not only sheds light on possible future developments but also provides general guidelines for policy formulation and for interpreting the findings 
within the Italian case.
First, we recall that approximately half of the respondents report that they do not believe their surrounding environment influences their health.
Although this result may appear surprising given the well-established evidence that environmental conditions affect health \citep[e.g.][]{pruss2016preventing}, it aligns with literature documenting limited health literacy in several European countries \citep{sorensen2015health}. Some studies specifically address the health-environment relationship \citep{MARTINKERRY202385, bert2023environmental,graham2022public}.
If this result reflects a broader lack of awareness regarding environmental influences,  explicitly highlighting this connection and relating it to general health literacy may help increase public awareness.
In particular, clarifying how the surrounding environment affects health could encourage healthier daily choices and prompt citizens to engage more proactively in improving their local environment.

Beyond this, our results reveal a strong territorial pattern in environmental perception. 
This emerges in Figure  

\ref{fig:regional_influence}, which reports the marginal regional proportions of the conditional responses, and it persists in Figures \ref{fig:plane_analysis2} and \ref{fig:asl_effects}, where marked local deviations remain even after accounting for both individual- and municipal-level covariates.
From a policy perspective, strong territoriality underscores the need for localized approaches that account for the specific characteristics of each area.
For instance, Italy has been facing depopulation in inner and rural areas (``aree interne''), with a gradual increase in the percentage of the population living in large cities and urbanized areas \citep{vendemmia2021institutional, benassi2023neighbourhood}.
 The observation that perceptions tend to worsen in highly populated and polluted areas, even after adjusting for multiple covariates, suggests that citizens in these contexts may be more sensitive to environmental issues.
 Thus, policies aimed at improving environmental quality in urban and densely populated areas may be both more readily accepted by these citizens and more effective for improving their health and perceived well-being.
 This does not imply that less urbanized or inner areas should receive less attention: these contexts face specific environmental challenges and may require equal or even greater effort. 
 In fact, while in urban settings the link between environmental quality and health may be more immediately recognized, in less urbanized areas raising awareness can be particularly challenging when a ``nice and clean'' landscape is deceptive and masks less visible forms of pollution, such as contaminated water (e.g., PFAS) or other hard-to-detect environmental stressors.


We recall that the variability explained by the considered covariates constitutes only a portion of the variability explained by the linear predictors.
Thus, we may hypothesize the presence of unconsidered factors that shape environmental perception, for which additional investigations are required in future studies.
This is particularly true for LHUs that deviate substantially from the main model, such as those highlighted in Section \ref{subsec:res}.
As noted, these are often areas with distinctive characteristics, and the analytical focus must account for their specific features.
While this may mean to include in the model other environmental concerns recognized in the literature and in specific territories (e.g., PFAS, NO2, asbestos, risk of flooding, etc.), this may also mean the necessity to reflect more on
the bidimensional nature of the response (to better recognize the determinants of positivity and neutrality), or to include in the model broader environmental settings shaped also by social and structural factors. 
A crucial aspect is to reflect on the question wording to mitigate potential multiple interpretations of the response options, a well-known challenging aspect in survey design \citep{Tourangeau_Rips_Rasinski_2000, groves2011survey}.


In addition to the concerns mentioned, it is worth exploring the inclusion of potential drivers of strictly positive responses, such as information on urban space (green areas), local environmental attitudes, or other socio-economic indicators, for both micro and macro areas not included in the presented model.


}


\section{Conclusions}\label{sec:conclusion}

A better understanding of people's perception of the potential environmental effects on personal health is fundamental for better targeting and evaluating public policies.
We believe that this study takes a further step in this direction,
by analyzing the data of the nationally representative PASSI surveillance system to understand the determinants that shape environmental perceptions in Italy. 
By integrating subjective survey responses with municipal-level contextual indicators, we highlight key aspects of how Italians perceive the influence of their environment on health, while also illustrating both the explanatory potential and the current limitations of the available data, pointing out even the need for further research.

In the first instance, we observe that a large share of the population exhibits a neutral or indifferent stance, but also that perceptions vary across the territory. 
Understanding this indifference is crucial from a policy and communication perspective, as targeted awareness campaigns could increase public engagement, ensuring not only the acceptance of environmental policies but also effective participation in initiatives requiring citizen involvement.
Methodologically, the semi-parallel penalized ordinal regression model is effective in handling complex ordinal data, providing a balance between predictive performance and interpretability. 
Representing the regression coefficients along two orthogonal dimensions--positivity and neutrality--offers new lens for interpretation. 
Ideally, we would increase positive perception--in response to actual environmental improvements--and reduce the beliefs of neutral attitudes, a key asset for fostering improvements in the surrounding area.
Our analysis reveals clear territorial patterns, including a persistent North–South gradient,
consistent with other observations on the health status and health determinants in Italy \citep[see, e.g.][]{minardi2011social}.
We find that negative perceptions tend to align with contextual risk factors, such as municipal-level PM2.5 concentrations and densely populated municipalities. 
These contextual indicators interact with individual factors that are associated with positive perceptions--such as higher education, good economic status, and living with other people--partially mitigating their positive effects, while also reducing the feeling of neutrality.

By contrast, when examining the results with respect to the response scale, it becomes clear that, beyond the strong territorial pattern, the covariates considered can explain only a small portion of the overall variability. 
This becomes particularly evident when examining the determinants of strictly positive responses.
On one side, this highlights the complexity of interpreting these perceptions and suggests that additional social, cultural, or psychological factors may be at play, warranting further investigation.
On the other side, this may also reflect intrinsic limitations of the survey question, where a neutral response signals the absence of recognized problems, which can be challenging to distinguish from a positive response, whereas a negative response clearly indicates recognized problems.
 
Despite some limitations discussed above, we believe that our findings advance both substantive and methodological understanding of environmental perceptions in Italy. 
They not only deepen knowledge of how individuals relate their environment to health but also provide actionable insights for policy by highlighting territorial disparities and the mechanisms that shape perceptions.
Future research should integrate richer data sources, conduct targeted territorial analyses, and extend the methodological framework to accommodate multiple survey items in order to capture the full complexity of environmental perceptions.

\section*{Acknowledgements} 
The authors are grateful to E. De Cian and to the members of the So.Sta. group (A. Arletti, M. Bertani, M. Marzulli, A. Pastore, M. Pittavino, L. Schiavon) at Ca' Foscari University of Venice for their insightful comments.
We also thank the Gruppo Tecnico PASSI (PASSI national coordination team at ISS) for their valuable feedback, and the PASSI network interviewers for their diligent work in data collection.

\section*{Data availability}
PASSI surveillance data can be accessed at http://www.epicentro.iss.it/passi/.
The dataset used for the analyses is not publicly available due to specific policies of the National Institute of Health and of the Italian Ministry of Health, but it is available from the National Institute of Health upon reasonable request.

\section*{Funding}
This project is developed under the project Planet4Health, which is funded by the European Commission grant 101136652. The five Horizon Europe projects, GO GREEN NEXT, MOSAIC, PLANET4HEALTH, SPRINGS, and TULIP, from the Planetary Health Cluster.

\bibliography{ref}

\end{document}